\documentclass[useAMS,usenatbib]{mn2e}

\usepackage[T1]{fontenc}
\usepackage{pslatex}
\usepackage{amsmath}
\usepackage{amsfonts}
\usepackage{color}
\usepackage{url}
\usepackage{graphicx}
\usepackage{subfigure}
\usepackage{amssymb}
\usepackage{textcomp}
\usepackage{soul}
\usepackage{booktabs}
\usepackage{array}
\usepackage{hyperref}
 \graphicspath{{images/}}

{\newif\ifnotend
\notendtrue
\def\veclist{ABCDEFGHIJKLMNOPQRSTUVWXYZabcdefghijklmnopqrstuvwxyz.}
\def\top#1#2.{#1}
\def\tail#1#2.{#2.}
\loop\expandafter\xdef\csname v\expandafter\top\veclist\endcsname%
{{\noexpand\bf\expandafter\top\veclist}}
\edef\veclist{\expandafter\tail\veclist}
\if\veclist.\notendfalse\fi\ifnotend\repeat}
\mathchardef\mhyphen="2D

\title[Converting sinks into stars]{A simple method to convert sink particles into stars}
\author[Sormani, Tre{\ss}, Klessen \& Glover]{Mattia C. Sormani$^1$, Robin G. Tre{\ss}$^1$, Ralf S. Klessen$^{1,2}$ and Simon C.O. Glover$^1$ \\
$^1$Institute for Theoretical Astrophysics, Zentrum f\"ur Astronomie der Universit\"at Heidelberg, Albert-\"Uberle-Str. 2, 69120 Heidelberg, Germany \\
$^2$Universit\"at Heidelberg, Interdiszipli\"ares Zentrum f\"ur Wissenschaftliches Rechnen, Im Neuenheimer Feld 205, D-69120 Heidelberg, Germany
}
\begin{document}

\date{}

\def\p{\partial}
\def\Omegap{\Omega_{\rm p}}

\newcommand{\di}{\mathrm{d}}
\newcommand{\bfx}{\mathbf{x}}
\newcommand{\bfe}{\mathbf{e}}
\newcommand{\vlos}{\mathrm{v}_{\rm los}}
\newcommand{\Tspin}{T_{\rm s}}
\newcommand{\Tb}{T_{\rm b}}
\newcommand{\degree}{\ensuremath{^\circ}}
\newcommand{\Th}{T_{\rm h}}
\newcommand{\Tc}{T_{\rm c}}
\newcommand{\bfr}{\mathbf{r}}
\newcommand{\bfv}{\mathbf{v}}
\newcommand{\pc}{\,{\rm pc}}
\newcommand{\kpc}{\,{\rm kpc}}
\newcommand{\Myr}{\,{\rm Myr}}
\newcommand{\Gyr}{\,{\rm Gyr}}
\newcommand{\kms}{\,{\rm km\, s^{-1}}}
\newcommand{\de}[2]{\frac{\partial #1}{\partial {#2}}}
\newcommand{\cs}{c_{\rm s}}
\newcommand{\rb}{r_{\rm b}}
\newcommand{\rqu}{r_{\rm q}}
\newcommand{\nuP}{\nu_{\rm P}}
\newcommand{\thetaobs}{\theta_{\rm obs}}
\newcommand{\hatn}{\hat{\textbf{n}}}

\maketitle

\begin{abstract}
Hydrodynamical simulations of star formation often do not possess the dynamic range needed to fully resolve the build-up of individual stars and star clusters, and thus have to resort to subgrid models. A popular way to do this is by introducing Lagrangian sink particles, which replace contracting high density regions at the point where the resolution limit is reached. A common problem then is how to assign fundamental stellar properties to sink particles, such as the distribution of stellar masses.

We present a new and simple statistical method to assign stellar contents to sink particles. Once the stellar content is specified, it can be used to determine a sink particle's radiative output, supernovae rate or other feedback parameters that may be required in the calculations.  Advantages of our method are (i) it is simple to implement, (ii) it guarantees that the obtained stellar populations are good samples of the initial mass function, (iii) it can easily deal with infalling mass accreted at later times, and (iv) it does not put restrictions on the sink particles' masses in order to be used. The method works very well for sink particles that represent large star clusters and for which the stellar mass function is well sampled, but can also handle the transition to sink particles that represent a small number of stars.
\end{abstract}


\section{Introduction}

Star formation involves the fragmentation of gaseous clouds, which then undergo local gravitational collapse. Following the global evolution of a cloud and, at the same time, the gravitational collapse of individual substructures is usually a task beyond the capabilities of modern numerical simulations \cite[e.g.][]{KlessenGlover2016}. 

A possible way to deal with this problem is the introduction of sub-resolution scale models for stellar birth. A very popular approach involves so-called sink particles.\footnote{sometimes also called star particles in situations where they are not allowed to accrete, usually in the context of cosmological simulations, e.g. \cite{Hopkins++2014,Hu++2016}.} This concept was first introduced by  \cite{Bate++1995}, and then used and developed further by many authors \cite[e.g.][and references therein]{Krumholz++2004,Jappsen++2005,Federrath++2010,Howard++2014,BleuerTeyssier2014,Klassen++2016,Gatto++2016}. In the sink particle approach, a contracting high-density region is replaced by a single Lagrangian particle at a stage where the numerical resolution limit in the simulation is reached. The particle inherits the mass as well as the linear and angular momentum of the original region, and in many implementations it can also accrete mass infalling at later times.

There are two regimes in which sink particles are used. If the resolution is high enough, each sink particle can represent a single star. If the resolution is poorer (for example because the simulated region is very large), then a single sink particle may correspond to an entire star cluster. In the latter case, a rule is needed to assign to each sink particle the appropriate stellar content. Several algorithms have been proposed to perform this task \citep[e.g.][]{Howard++2014,Dale++2014,Gatto++2016,Hu++2016}. However, they can usually be employed only if the sink particle mass distribution satisfies some restrictions. Typically they require a minimum mass to be exceeded, both at the time of formation and during the subsequent accretion phase. Moreover, even though they rely on stochastic sampling from the stellar initial mass function \citep[IMF: see][]{Kroupa2002, Chabrier2003} they can introduce biases which prevent the final result from being a completely faithful representation of the IMF.

In this short paper, we propose a new statistical method to assign a realistic stellar content to cluster sink particles. Once the distribution of stellar masses and formation times are determined, this information can be used to assign the radiative output of the sink particle, its supernovae rate, chemical yields, and so forth. The method is introduced in Section \ref{sec:method} and an illustrative example is presented in Section \ref{sec:Kroupa}. We summarise and conclude in Section \ref{sec:conclusion}.

\section{The method} \label{sec:method}

\subsection{Definition}

Consider an IMF, with $N$ different types of stars binned according to their masses. This defines a discrete set of stellar masses  
\begin{equation}
\{m_1, m_2, \ldots, m_N\} \,,
\end{equation}
with corresponding mass fractions
\begin{equation}
\{f_1, f_2, \ldots, f_N\}\, \quad \mathrm{such\; that} \quad \sum_i f_i = 1 \, ,
\end{equation}
i.e., in a well-sampled population a fraction $f_i$ of the total mass will be in stars of type $i$. The statistical weights $f_i$ can be easily determined starting from more common ways of parametrising the IMF (see Section \ref{sec:Kroupa}) and depend on the width of the chosen bins.

Given a sink of mass $M$, we now describe our procedure to assign stars to it. Let us denote the stellar content by a vector 
\begin{equation}
\mathbf{n} = \{n_1, n_2, \ldots, n_N\}  \,,
\end{equation}
 where the $n_i$'s are integers representing the number of stars of type $i$. So for example a sink with no stars has $\mathbf{n}=\{n_1=0, n_2=0, \ldots \}$ and a sink with just 2 stars of type 1 has $\mathbf{n}=\{n_1=2, n_2=0, n_3=0, \ldots \}$. Let us call $P(\mathbf{n})$ the probability that the stellar content assigned to the sink particle is $\mathbf{n}= \{n_1, n_2, \ldots \}$. Our prescription is:

\begin{equation}
P(\mathbf{n}) = P_1(n_1) P_2(n_2) \ldots P_N(n_N)\, ,\label{eq:def1}
\end{equation}
where
\begin{equation}
P_i(n_i)=e^{-\lambda_i}\frac{\lambda_i^{n_i}}{n_i !}\, , \label{eq:poisson}
\end{equation}
and
\begin{equation}
\lambda_i = f_i \frac{M}{m_i}\,. \label{eq:lambdai}
\end{equation}
In other words, the number of stars of each type are assigned according to a Poisson distribution with mean $\lambda_i$ which depends on the mass of the sink particle. The total stellar mass assigned to the sink will then be equal to 
\begin{equation}
M_{\star} = n_1 m_1 + n_2 m_2 + \ldots + n_N m_N\,,
\end{equation}
and the stellar mass in stars of type $i$ will be equal to:
\begin{equation}
M_{\star, i} = n_i m_i\,.
\end{equation}

\subsection{Properties} \label{sec:properties}

The method defined in the previous section has the following properties (see Appendix \ref{sec:appendix} for a derivation of these properties):
\begin{enumerate}
\item The total stellar mass is on average equal to the sink mass $M$:
\begin{align}
	\langle M_{\star} \rangle 	& =  \sum_{\mathbf{n}} M_{\star} P(\mathbf{n}) 
						 = M \, .
\end{align}
Here and in the following $\langle . \rangle$ denotes the ensemble average over many random realisations of the Poisson process and $\sum_{\mathbf{n}}~\equiv~\sum_{n_1,n_2,\ldots,n_N}.$
\item The variance of the total stellar mass $M_\star$ is:
\begin{align}
\frac{ \langle \left( M_\star - M \right)^2 \rangle}{M^2}  & =  \sum_{\mathbf{n}}  \frac{ (M_\star - M)^2}{M^2} P(\mathbf{n})  
										 = \frac{\bar{m}}{M} \, , \label{eq:varMtot}
\end{align}
where
\begin{equation}
\bar{m} = \sum_i f_i m_i
\end{equation}
is the mean stellar mass according to the IMF. 
\item The average mass in stars of type $i$ is:
\begin{align}
{ \langle M_{\star, i} \rangle }					& = \sum_{\mathbf{n}} m_i n_i  P(\mathbf{n})
											 = f_i M \,.
\end{align}
\item The variance of the mass in stars of type $i$ is:
\begin{align}
\alpha_i^2 \equiv \frac{ \langle  \left( M_{\star, i}  - f_i M \right)^2   \rangle }{(f_i M)^2} 	& = \sum_{\mathbf{n}} \frac{ \left( m_i   n_i  - f_i M \right)^2 }{(f_i M)^2}   P(\mathbf{n}) 
													= \frac{m_i}{f_i M} \, . \label{eq:varMi}
\end{align}

\item The stellar content $\mathbf{n}$ assigned to a sink particle of mass $M~=~M_1~+~M_2$ is statistically equivalent to the sum $\tilde{\mathbf{n}}=\mathbf{n}_1+\mathbf{n}_2$ of the stellar contents assigned to sink particles of masses $M_1$ and $M_2$. In other words, $\mathbf{n}$ and $\tilde{\mathbf{n}}$ have exactly the same probability distribution. This is a direct consequence of the additive property of the Poissonian process, i.e. the sum of two Poisson distributions is a Poisson distribution whose mean is the sum of the means of the original distributions \citep[e.g.][]{haight1967handbook}.

\end{enumerate}

\subsection{Discussion of the properties} \label{sec:discuss}

Since the method is stochastic in nature, a sink particle mass $M$ and its stellar mass $M_\star$ will in general differ. However property (i) ensures that given many realisations of the process the two masses are on average the same. The typical deviations of the stellar mass $M_\star$ from the sink particle mass $M$ are quantified by property (ii): they will be small if $M \gg \bar{m}$, which for typical applications is $\bar{m}~\sim~\text{few solar masses}$.

Property (iii) demonstrates that the IMF is always recovered on average. Property (iv) quantifies the deviation in the number of stars of each type obtained from the average number that would be expected according to the IMF. From equation \eqref{eq:varMi} we see that the smaller the $f_i$'s (the rarer the star), the larger the deviations. The precise values of the $f_i$'s depend on how one chooses to discretise a continuous IMF by binning it. Broad bins will lead to smaller deviations, narrow bins will result in larger fluctuations. How one should choose the bin sizes depends on the fluctuations that can be tolerated and on the science question to be addressed. The method also works well with bins of unequal sizes. For example, if one is only interested in the number of high-mass stars, it may be appropriate to use fine binning in the high-mass regime, and combine all low-mass stars into one single bin. The stochastic fluctuations in each bin $i$ depend on the sink particle mass and the bin 
 width and can be calculated from equation \eqref{eq:varMi}. We discuss this further in Sect. \ref{sec:Kroupa}.

Property (v) has nice consequences. First, accretion onto the sink particle is naturally taken care of. If a sink particle of mass $M$ later accretes a mass $\Delta M$, the extra mass can be converted into stars and the result would be the same as if all the original plus the accreted mass is converted at once. Thus if the same sink particle accretes mass in many stages, each time we can convert the mass into stars without worrying that the final distribution of stars will be affected by when we do it. Moreover, if the sampling is done with sufficiently fine time stepping, the model will automatically produce a time-distribution for the birth of stars which can then be used in the scientific investigation at hand. If a sink with initial mass $M\gg\bar{m}$ later accretes only a small quantity of mass (e.g., comparable to the mass of a star), then our method still works well. The extra mass will produce few, one or no stars, which will be assigned to the sink in addition to the stars assigned initially, when the sink was created. At this later time, the total stellar content of the sink will be a better sample of the IMF than it was before. So even if we accrete a small amount of mass, the overall content of the sink always tends towards a better representation of the IMF, never worse. 

Second, the only statistically important parameter for a collection of sink particles is their total mass $M$. The total distribution of stars is independent of how these masses are distributed across the sinks. This property makes our approach quite insensitive to the numerical resolution, because it does not matter if the mass of the emerging star cluster is contained within one single massive sink (in a lower resolution simulation) or distributed over particles of smaller masses (as would form in the same region in simulations with better spatial resolution). The overall distributions will be exactly the same. 

The fact that the stellar mass and the sink mass can be different may create problems in certain circumstances where a careful conservation of mass and momentum is needed. The simplest way to address this issue is to let each particle have two masses: a dynamical mass ($M$) and a stellar mass ($M_\star$). The code will continue to use $M$ as the dynamical mass, while $M_\star$ is used only to calculate radiative output and stellar-related properties of the sink. In this way, the stellar content is just a label attached to the sink.

As discussed above, fluctuations will be small when $M \gg \bar{m}$. In the intermediate regime where, say, $M \simeq 100\,M_\odot$, property (ii) guarantees that fluctuations in the total mass will still be small. When approaching the low mass regime, in which the mass of an individual sink is comparable to the mass of an individual star, $M \simeq \bar{m}$, our scheme can still be used but one needs to evaluate the situation more carefully. In this regime, our scheme will usually assign few, one or no stars at all to each sink. Clearly, the same happens in nature for small clusters which contain only few stars. Hence, our scheme automatically covers the fluctuations of the IMF that we do see in real clusters of small mass. Obviously both in our scheme and in nature the stellar content of an individual sink or small cluster containing very few stars cannot be a good sample of the IMF. Even if the stellar content of individual sinks can fluctuate wildly, when we consider many small clusters of small mass together, we eventually recover the IMF (thanks to property (v) above), exactly as it would happen in nature if we did a statistical study by considering many small clusters together.

However our scheme has a shortcoming and allows for something that nature forbids: the stellar mass $M_\star$ assigned to a sink can be larger than the sink mass $M$. This effect is much more severe in the regime of small sink masses (see equation \ref{eq:varMtot}). Whether this is a problem in practice depends on the aims of the scientific investigation at hand. For example, in feedback simulations in which one is interested in producing a realistic population of supernovae and analysing their effect on the surrounding ISM, our scheme works well. It can produce a realistic distribution of supernovae, both in space and in time. Our method works less well if we are interested in modelling the evolution of one particular cluster (rather than the average effect of many clusters on the ISM through their stellar feedback), as in this case we may assign to the sink of interest a stellar mass $M_\star$ which is too far off from its mass $M$.

Finally, we note that depending on the underlying physical model not all the mass in the sink particle needs to be converted into stars, i.e. the star formation efficiency may not be $100 \%$. In this case, we should modify our procedure by replacing $M$ by an effective mass $M_{\rm eff}$ in the definition of $\lambda_i$ in equation \eqref{eq:lambdai}. For example, if the star formation efficiency is, say, $50 \%$, then we can use $M_{\rm eff}= 0.5 M$.

\subsection{Comparison with other methods} \label{sec:others}
 
Other methods that have been proposed \citep[e.g.][]{Howard++2014,Gatto++2016} usually involve a procedure along the following lines. First, a gas reservoir is associated to the sink particle. The reservoir is then gradually converted into stars by sampling directly from the IMF. At each step of the sampling process, the gas reservoir diminishes. The main problem with such methods is to decide at which point (i.e., for which reservoir mass) one should stop sampling. For example, when the gas reservoir is too small, sampling from the IMF can lead to a star whose mass exceeds that available in the reservoir. What decision should be made at this point?

There are various possibilities. \cite{Howard++2014} accepts the outcome of the sampling only if the mass of the star does not exceed the mass of the reservoir. However, this introduces a bias and the IMF will be oversampled at the low-mass end. To circumvent this problem, these authors assumed that accretion is present at a certain rate and devise a procedure for fine-tuning the sampling time, and have shown that under such conditions the bias is negligible. However this introduces restrictions on the sink particles mass distribution in order to work and depends on the details of the accretion scheme and in general on the details of the simulations.

Another possibility is to stop the sampling process when the gas reservoir is too small. However this method cannot handle sink particles whose mass is below that of the most massive star. \cite{Gatto++2016} and \cite{HennebelleIffrig2014} follow an approach along similar lines. These authors are mostly concerned about the high-mass end of the IMF ($M > 8 \,{M_\odot}$), which is needed to model stellar feedback in the ISM. They randomly draw one massive star for every $120 \,{M_\odot}$ of accreted mass. The difference between this value and the sampled mass is assumed to go into low-mass stars. Since the mass of each sink is not in general an exact multiple of $120 \,{M_\odot}$, this produces an overall deficit of high mass stars and the procedure has to be modified by considering the mass of all the sinks together \citep{Kortgen++2016}. The stars so created have then to be redistributed to the sinks according to some criteria.

Simpler methods that do not involve statistical sampling from the IMF have also been proposed to assign specific properties to sink particles. \cite{Dale++2014}, for example, compute the mass fraction in high mass stars but then turn on the corresponding feedback  only if the sink mass is above a certain threshold value. \cite{Hopkins++2014} and \cite{Hu++2016} assign stochastically discrete forms of feedback, such as supernovae, to the sinks based on their mass, age and metallicity. The limitations of these models are linked to the lack of a full stellar content of the sink, and so the properties inferred this way can only be coarse-grained representations of the true stellar population. This may be appropriate when modelling stellar feedback in cosmological galaxy formation calculations, where individual sink particles are very massive and can contain multiple clusters, but it is not sufficient in high resolution simulations where sink particles represent small stellar clusters or in situations where a more detailed modelling of stellar evolution at the sub-grid level is needed.

Our method bypasses most of the problems mentioned above. It  guarantees that the IMF is well sampled without putting any restriction on the sink particle masses or on the sampling time. It also makes it easy to control the average stellar mass produced, so that unwanted overall deficits or surpluses are avoided. It naturally gives the full stellar content, including the stars of each type, and not only the number of high mass stars. Moreover, our method is conceptually simple and computationally efficient since it only needs to draw from independent Poisson distributions for which standard routines are available in all programming languages. 

As discussed in Section \ref{sec:discuss}, a possible shortcoming of our method is that it is in principle possible to assign to a sink a stellar mass $M_\star$ that exceeds the mass of the sink $M$. However, this in unlikely to be a problem in most applications, particularly when the star formation efficiency is less than $100\%$ and $M\gg \bar{m} \simeq \text{few } M_\odot$, in which case the variance can also be seen as a fluctuation in the star formation efficiency. As mentioned in Sect. \ref{sec:discuss}, we suggest treating the stellar content as a mere label attached to the particle, and keeping the sink particle mass as the dynamical mass in the underlying hydrodynamic simulation. 

\section{Example with a Kroupa IMF} \label{sec:Kroupa}

In this section we briefly apply our method with the IMF as parametrised by \cite{Kroupa2002}. In this approach, the number of stars in the mass interval $(m,\, m+\di m)$ is given by $f(m)\di m$ with
\begin{equation}
f(m) = 
\begin{cases} 
A m^{-0.3} \quad \text{for} \quad \tilde{m}_1 \leq m < \tilde{m}_2 \\
A k_1 m^{-1.3} \quad \text{for} \quad \tilde{m}_2 \leq m < \tilde{m}_3 \\
A k_2 m^{-2.3} \quad \text{for} \quad \tilde{m}_3 \leq m 
\end{cases}\, ,
\end{equation}
and where $A$ is a global normalization factor, $k_1=\tilde{m}_1^{-0.3+1.3}$, $k_2=k_1 \tilde{m}_2^{-1.3+2.3}$, $\tilde{m}_1 = 0.01\, M_\odot$, $\tilde{m}_2 = 0.08\, M_\odot$ and  $\tilde{m}_3 = 0.5 \,M_\odot$.

We have binned Kroupa's IMF in mass bins in the range $M_{\rm min}=0.01\,M_\odot$ to $M_{\rm max}= 100\,M_\odot$. The $m_i$'s and $f_i$'s are calculated according to the following formulae:
\begin{align}
& m_i =  \frac{ \int_{i} m f(m)\di m}{\int_{i} f(m) \di m}, \label{eq:mi} \\
& f_i =  \frac{ \int_{i} m f(m)\di m}{\int_{M_{\rm min}}^{M_{\rm max}} m f(m) \di m}, \label{eq:fi}
\end{align}
where $\int_{i}$ denotes the integral over the $i$-th bin.  

\begin{figure}
\includegraphics[width=0.5\textwidth]{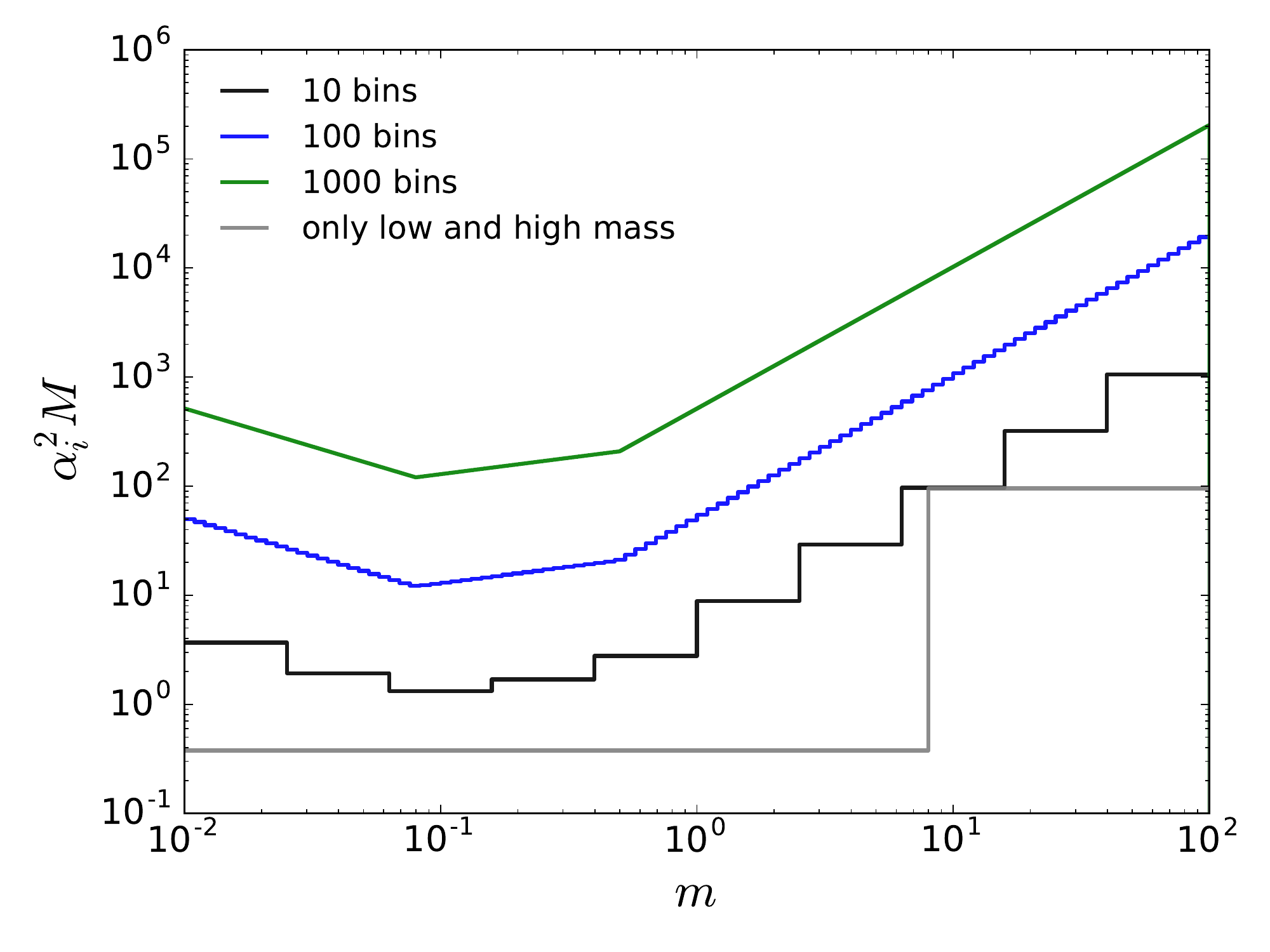}
 \caption{The values of $\alpha_i^2 M$, which quantifies the fluctuations in the number of stars in each bin, for four different binnings of the IMF in the range $0.01\,M_\odot \mhyphen 100\,M_\odot$. The first three lines from the top are for logarithmic binnings with 1000, 100 or 10 bins respectively. The bottom line is for a binning that distinguishes only between low mass ($<8\,M_\odot$) and high mass ($>8\,M_\odot$) stars.}
\label{fig:bins} 
\end{figure}

Figure \ref{fig:bins} shows the value of $\alpha_i^2 M$ for four different ways of binning. The quantity $\alpha_i$ is the expected error in the number of stars assigned to each bin, i.e., it is the deviation from the number actually assigned in a random realisation and the average number predicted by the IMF (see equation \ref{eq:varMi}). We can use Fig. \ref{fig:bins} to read off directly the minimum mass $M$ needed for the error $\alpha_i$ to be below a given tolerance value. The three top lines in Fig. \ref{fig:bins} refer to logarithmic binnings with 1000, 100 or 10 bins, respectively. We see that, as one would expect, for a particle of given mass $M$ fine binning leads to larger fluctuations. This means that if we want to distinguish between many types of stars and still get a well sampled IMF, we need sink particles with a large mass that can contain many stars (the same is true in nature for real clusters: we cannot have a good sample of the IMF looking at a too small cluster). The gray line at the bottom is for a very coarse-grained binning that distinguishes only between low mass ($<8M_\odot$) and high mass ($>8M_\odot$) stars, and it shows that in this case lower 
 values of $M$ are permitted to achieve a given tolerance target.

\begin{figure*}
\includegraphics[width=1.0\textwidth]{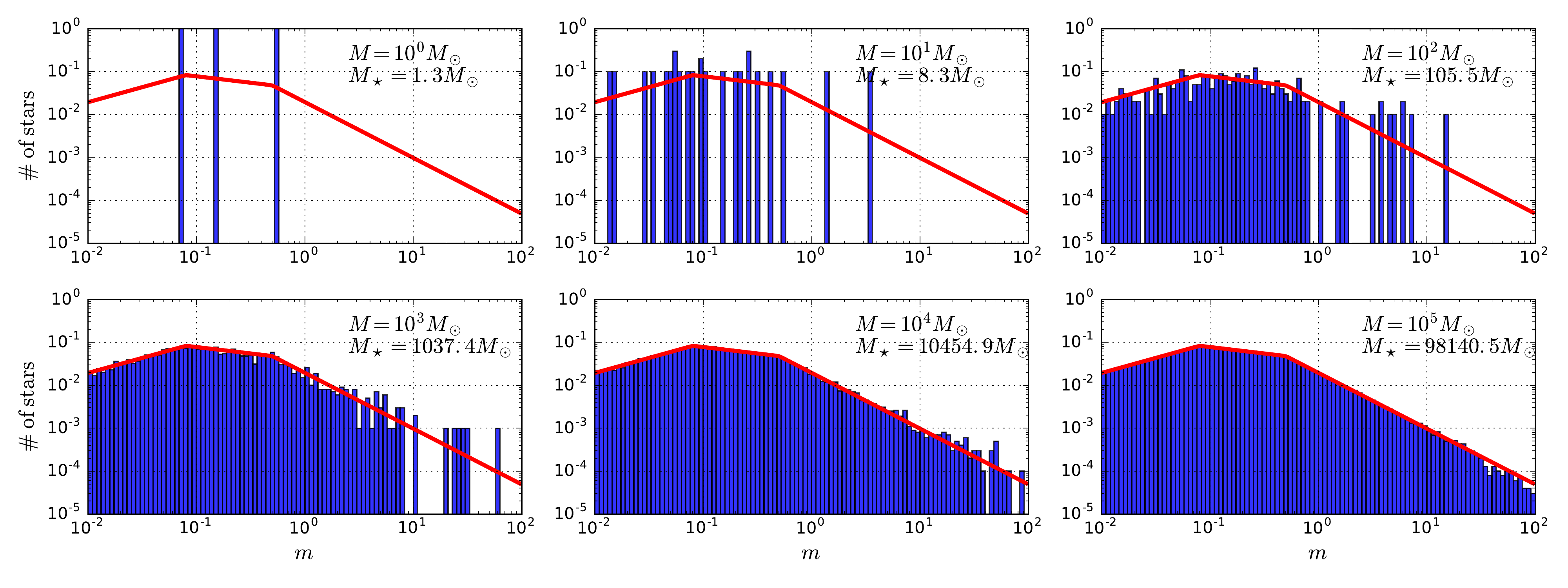}
 \caption{Application of our method using Kroupa's IMF. Different panels show random realisations for different values of the mass $M$. The blue bars show the number of stars in each mass bin, normalised with arbitrary scaling. The red line shows the number of stars in each bin that would be expected according to Kroupa's IMF.}
\label{fig:Kroupa} 
\end{figure*}

Each panel in Fig. \ref{fig:Kroupa} shows a random realisation calculated according to our method for a sink particle of given mass $M$ and a binning with 100 logarithmic bins (blue line in Fig. \ref{fig:bins}). Different panels differ only in the value of $M$. It is clear that when $M = 1\ M_\odot$ the IMF is not well sampled. As $M$ is increased, the lower-mass end of the IMF gets well sampled first, while the higher-mass region retains larger sampling uncertainties, as one would expect from the steeply decreasing probability of forming high-mass stars. Figure \ref{fig:bins} also explains this behaviour. For the case of 100 logarithmic bins, the blue line indicates that the fluctuations at the low-mass end are small (say $\alpha_i \lesssim 10\%$)  for $M \gtrsim \text{few } 10^3\,M_\odot$, and at the high-mass end are small when $M \gtrsim \text{few } 10^4\,M_\odot$, which is exactly the behaviour seen in Fig. \ref{fig:Kroupa}. If we can get by with fewer mass bins, we can reach the same
  fluctuation level $\alpha_i$ with lower-mass sinks. In many applications, the decision on how much variance $\alpha_i$ can be tolerated at a certain mass $m_i$ determines the choice of the bin size $f_i$, which can then be calculated from equation (\ref{eq:varMi}) for a given mass $M$. Keeping in mind property (v), we can interprete Figs. \ref{fig:Kroupa} and \ref{fig:bins} either as referring to the mass of a population of sink particles summing up to a total of $M$ or as referring to one single sink particle with mass $M$. This means we have the choice of whether we apply the maximum tolerance to a single sink particle or to the population as a whole.

We note that in this example we have focussed on the \cite{Kroupa2002} IMF and discretised it in four different ways. However, the method can be applied with no extra complication to different IMF models and to bins of any sizes and non-uniform width. All one needs to do is calculate the appropriate values of the $f_i$'s and $m_i$'s for the desired IMF and mass binning. The method can be also generalised to the continuous limit if one wants to avoid binning (see Appendix \ref{sec:continuous}).

\section{Conclusion} \label{sec:conclusion}

We have presented a simple prescription for assigning stellar contents to sink particles used in hydrodynamical simulations with star formation that do not fully resolve the build-up of individual stars and star clusters. The assigned stars can then be used to determine the sink particle's radiative output, its supernova rate and chemical yield, or other parameters of interest related to the underlying stellar population. The key properties of our method are summarised as follows:
\begin{itemize}
\item It guarantees that the assigned stellar populations are a faithful representation of the IMF.
\item It can  easily deal with infalling mass accreted at later times by the sink particle.
\item  It does not put any restriction on the sink particle masses or on the sampling time. For example, there is no minimum mass requirement for the sink particles. The method therefore is relatively insensitive to the numerical resolution adopted in the simulation. 
\item It can accomodate in a straightforward way non-uniform mass bins and arbitrary choices of the IMF.
\item It can deal with sinks of intermediate masses, i.e. in the intermediate regime between the cluster regime and the star regime, thus providing a smooth transition between the two.
\item In the intermediate-small sink mass regime, it can cover the statistical fluctuations seen in nature in real clusters of small size.
\item It is computationally very efficient and easy to implement.
\end{itemize}
Some possible shortcomings are:
\begin{itemize}
\item Due to the stochastic nature of the method, the sink particle mass $M$ will be equal to its stellar mass $M_\star$ only on average, but not in every single instance. This can cause problems when approaching a physical regime where sink particles describe few or single stars. Our method works best when $M \gg \bar{m}$ as this shortcoming becomes negligible in this limit.

\item The IMF is taken as input to the model. It is independent of the sink particles' mass distribution from the numerical simulation. That means that this quantity cannot be used as a diagnostic tool.
\end{itemize}

\section*{Acknowledgements}

The authors thank Paul Clark, Samuel Geen, Diederik Kruijssen, Mordecai Mac Low, John Magorrian, Eric Pellegrini, Rowan Smith and all the members of the Heidelberg Institute for Theoretical Astrophysics for useful comments. We also acknowledge support from the Deutsche Forschungsgemeinschaft in the Collaborative Research Center (SFB 881) ``The Milky Way System'' (subprojects B1, B2, and B8) and in the Priority Program SPP 1573 ``Physics of the Interstellar Medium'' (grant numbers KL 1358/18.1, KL 1358/19.2). RSK furthermore thanks  the European Research Council for funding in  the ERC Advanced Grant STARLIGHT (project number 339177).

\appendix 

\section{Derivation of the properties} \label{sec:appendix}
In this appendix we prove the properties listed in Section \ref{sec:properties}. It is well known that the Poisson distribution given by equation \eqref{eq:poisson} has the following properties:

\begin{align}
 \sum_{\mathbf{n}} P(\mathbf{n}) & = 1, \label{eq:p1} \\ 
  \sum_{\mathbf{n}} n_i P(\mathbf{n}) & = \lambda_i, \label{eq:p2} \\ 
  \sum_{\mathbf{n}} n_i^2 P(\mathbf{n}) &= \lambda_i + \lambda_i^2. \label{eq:p3}
\end{align}
Properties (i) and (iii) in Section \ref{sec:properties} follow immediately from equations \eqref{eq:p1}, \eqref{eq:p2} and \eqref{eq:p3}. Property (ii) can be proved as follows

\begin{align}
\frac{ \langle \left( M_\star - M \right)^2 \rangle}{M^2} 
										& =  \frac{1}{M^2} \sum_{\mathbf{n}} M_\star^2 P(\mathbf{n}) - 1 \\
										& =  \frac{1}{M^2} \sum_{\mathbf{n}} (n_1 m_1 + \ldots + n_N m_N)^2 P(\mathbf{n}) - 1 \\
										& = \frac{1}{M^2} \left[ \sum_i \sum_{\mathbf{n}} n_i^2 m_i^2 P(\mathbf{n}) + 2 \sum_{i < j} \sum_{\mathbf{n}} n_i n_j m_i m_j P(\mathbf{n})\right]- 1 \\
										& = \frac{1}{M^2}  \left[ \sum_i m_i^2 (\lambda_i + \lambda_i^2) + 2 \sum_{i < j} m_i m_j \lambda_i \lambda_j \right]- 1 \\
										& = \frac{1}{M^2}  \left[ \sum_i m_i f_i M + f_i^2 M^2 + 2 \sum_{i < j} f_i f_j M^2 \right]- 1 \\
										& = \frac{1}{M^2}  \left[ \sum_i m_i f_i M + \left( \sum_i f_i \right)^2  M^2  \right]- 1 \\
										& = \frac{1}{M^2}  \left[ \sum_i m_i f_i M +  M^2  \right]- 1 \\
										& = \frac{\bar{m}}{M}
\end{align}
which is equation \eqref{eq:varMtot}. Property (iv) can be proved as follows:
\begin{align}
\alpha_i^2 = 	& \frac{ \langle  \left( M_{\star, i}  - f_i M \right)^2   \rangle }{(f_i M)^2} 	\\
		   	& = \sum_{\mathbf{n}} \frac{ \left( m_i   n_i  - f_i M \right)^2 }{(f_i M)^2}   P(\mathbf{n}) \\
			& = \sum_{\mathbf{n}} \frac{  m_i^2   n_i^2  + f_i^2 M^2 - 2 m_i n_i f_i M}{(f_i M)^2}   P(\mathbf{n}) \\
			& = \frac{  m_i^2  \left( \lambda_i + \lambda_i^2 \right)  + f_i^2 M^2 - 2 m_i \lambda_i f_i M}{(f_i M)^2} \\
			& = \frac{  m_i f_i M + f_i^2 M^2  + f_i^2 M^2 - 2 f_i^2 M^2}{(f_i M)^2} \\
			& = \frac{m_i}{f_i M}
\end{align}
which is equation \eqref{eq:varMi}. Property (v) follows immediately from the additive property of Poisson distributions quoted in the main text. 

\section{Continuous Limit} \label{sec:continuous}
Our method involves discretisation of the IMF, which may not be ideal in certain applications where one wants to avoid binning and deal directly with a continuous IMF. In the continuous limit, when the number of bins tends to infinity and the width of each bin tends to zero, our method gives results that are equivalent to the following two-step procedure. Given a sink particle of mass $M$:
\begin{enumerate}
\item Draw the number of stars to assign to the sink according to a Poisson distribution with mean $\lambda = M / \langle m \rangle $.
\item For each star, draw its mass directly from the (continuous) IMF.
\end{enumerate}
Only for this appendix, we have defined 
\begin{equation}
 \langle { m } \rangle = \frac{ \int m f(m) \di m}{\int f(m)\di m},
\end{equation} 
where $f(m)$ is the IMF (see Section \ref{sec:Kroupa}).

\def\aap{A\&A}\def\aj{AJ}\def\apj{ApJ}\def\mnras{MNRAS}\def\araa{ARA\&A}\def\aapr{Astronomy \&
 Astrophysics Review}\def\apjs{ApJS}\def\apjl{ApJ}\def\pasj{PASJ}\def\nat{Nature}\def\prd{Phys. Rev. D}
\def\ssr{Space Sci. Rev.}\def\pasp{PASP}
\bibliographystyle{mn2e}
\bibliography{bibliography}

\end{document}